\documentclass[aps,prd,twocolumn,superscriptaddress,showpacs,amsmath,amssymb,nofootinbib, preprintnumbers]{revtex4-1}
 \usepackage{amsmath}
\usepackage{graphicx}
\usepackage{epstopdf}
\usepackage{float}
\usepackage{color}
\usepackage[toc,page]{appendix}
\usepackage[usenames,dvipsnames]{xcolor}
\usepackage[normalem]{ulem}

\newcommand{\be}{\begin{equation}}
\newcommand{\ee}{\end{equation}}
\newcommand{\ba}{\begin{eqnarray}}
\newcommand{\ea}{\end{eqnarray}}

\newcommand{\beq}{\begin{equation}}
\newcommand{\eeq}{\end{equation}}
\newcommand{\beqa}{\begin{eqnarray}}
\newcommand{\eeqa}{\end{eqnarray}}

\begin{document}
\title{The First Law for Rotating NUTs}

\author{Alvaro Ballon Bordo}
\email{aballonbordo@perimeterinstitute.ca}
\affiliation{Perimeter Institute, 31 Caroline Street North, Waterloo, ON, N2L 2Y5, Canada}
\affiliation{Department of Physics and Astronomy, University of Waterloo,
Waterloo, Ontario, Canada, N2L 3G1}

\author{Finnian Gray}
\email{fgray@perimeterinstitute.ca}
\affiliation{Perimeter Institute, 31 Caroline Street North, Waterloo, ON, N2L 2Y5, Canada}
\affiliation{Department of Physics and Astronomy, University of Waterloo,
Waterloo, Ontario, Canada, N2L 3G1}

\author{Robie A. Hennigar}
\email{rhennigar@mun.ca}
\affiliation{Department of Mathematics and Statistics, Memorial University of Newfoundland, St. John’s, Newfoundland and Labrador, A1C 5S7, Canada}

\author{David Kubiz\v n\'ak}
\email{dkubiznak@perimeterinstitute.ca}
\affiliation{Perimeter Institute, 31 Caroline Street North, Waterloo, ON, N2L 2Y5, Canada}
\affiliation{Department of Physics and Astronomy, University of Waterloo,
Waterloo, Ontario, Canada, N2L 3G1}

\date{June 1, 2019}

\begin{abstract}

We address a long-standing problem of describing the thermodynamics of rotating Taub--NUT solutions. The obtained first law is of full cohomogeneity and allows
for asymmetric distributions of Misner strings as well as their potential variable strengths---encoded in the gravitational Misner charges.
Notably, the angular momentum is no longer given by the Noether charge over the sphere at
infinity and picks up non-trivial contributions from Misner strings.
\end{abstract}

\maketitle

The Lorentzian Taub--NUT spacetime is one of the simplest yet one of the most puzzling vacuum solutions of general relativity. The spacetime possesses a Schwarzschild-like horizon that, as a consequence of the NUT charge, is accompanied by rotating string-like singularities on the north/south pole axes, known as Misner strings.

The Misner strings can be exiled from the spacetime at the cost of introducing a periodic time coordinate~\cite{misner1963flatter}. However, while this approach is customary it is not strictly necessary: as suggested by Bonnor~\cite{bonnor1969new}, see also~\cite{Manko:2005nm}, the Misner strings can instead be interpreted as due to `singular' sources of angular momentum, and the geometry from this perspective is less pathological than one might expect. For example, the spacetime is geodesically complete and while closed time-like curves exist near the axes, there are no causal pathologies for geodesic observers  \cite{miller1971taub, Clement:2015cxa, Clement:2015aka, Clement:2016mll}. This raises a possibility that the NUT charge may actually be relevant for astrophysics. In fact, already in 1997 astrophysicists probed the possibility of detecting the NUT parameter by microlensing  \cite{NouriZonoz:1998va, lynden1998classical}, while its impact on the black hole shadow has been investigated in \cite{Abdujabbarov:2012bn, Grenzebach:2014fha, Goddi:2017pfy}. Ongoing and forthcoming tests of general relativity in the strong field regime, e.g. the Event Horizon Telescope \cite{Akiyama:2019cqa}, may reveal interesting signatures of these spacetimes or at the very least provide tighter constraints.

Notwithstanding their potential pathologies, solutions with NUT charge have been the source of deep physical insights, for example in the context of black hole thermodynamics. The thermodynamics of these solutions has been predominantly studied in the Euclidean case and in the absence of Misner strings, e.g. \cite{Hawking:1998ct, Chamblin:1998pz, Emparan:1999pm, Mann:1999pc, Mann:1999bt, Garfinkle:2000ms, Johnson:2014xza, Johnson:2014pwa}.  The most striking result to come from these studies is that, in equilibrium, the entropy of Taub--NUT solutions is not simply one quarter the horizon area, providing a counter-example to the `area law'.

A full understanding of the properties---in particular the first law---of the Lorentzian Taub--NUT solutions with Misner strings present is much subtler. In fact, to the best of our knowledge, until recently no consistent first law for these solutions has been presented. It was argued in~\cite{Astefanesei:2004ji, Mann:2004mi} that an additional relationship between the horizon radius and the NUT charge must be obeyed in order for the first law to hold. In~\cite{Holzegel:2006gn} a Smarr relation was derived for the Lorentzian solution, but no first law consistent with the Smarr formula was obtained. The situation was remedied for the non-rotating solution in~\cite{Kubiznak:2019yiu, Ballon:2019uha}, where it was found that these issues can be ameliorated through a judicious choice of a potential and conjugate charge for the NUT parameter.\footnote{See also~\cite{Bueno:2018uoy} where it was found that for Euclidean Taub--NUT solutions with toroidal bases (which do not feature Misner strings) the analogous terms are required for consistency of the first law.}

The purpose of this note is to address the first law for Taub--NUT solutions with rotation. The Euclidean setup was considered in~\cite{Ghezelbash:2007kw}, where it was found that regularity (equilibrium) of the Euclidean sector effectively leads to a `discrete' version of the first law. Here, focusing on the Lorentzian case and picking up the threads on \cite{Bordo:2019tyh}, our first law not only allows for an arbitrary distribution of Misner strings, it also accounts for their variable strengths.
The Misner gravitational charges that appear in this law are obtained by a (rotating-case) generalization of the Komar-like integrals introduced in \cite{Bordo:2019tyh} and
the results are cross-checked by the Euclidean action calculation. One of the key results is the realization that Misner strings provide non-trivial contributions to the angular momentum of the spacetime, which is no longer simply given by the Noether charge over the sphere at infinity.

Let us begin our exploration by introducing the rotating Taub--NUT spacetime. The solution reads
\ba\label{NUT}
ds^2&=&-\frac{\Delta}{\Sigma}\Bigl[dt+(2n\cos \theta+2s-a\sin^2\!\theta)d\phi\Bigr]^2
+\frac{\Sigma}{\Delta}\,dr^2\nonumber\\
&&+\frac{\sin^2\!\theta}{\Sigma}\Bigl[adt-(r^2+a^2+n^2-2as)d\phi\Bigr]^2\!\!+{\Sigma}d\theta^2\;,\!\!
\nonumber\\
\Delta&=&r^2+a^2-2mr-n^2\,,\nonumber\\
\Sigma&=& r^2+(n+a\cos\theta)^2\;,
\ea
where $n$, $m$, and $a$ are the standard NUT, mass, and rotation parameters, respectively. There is yet another dimensionful physical parameter $s$, which governs the overall distribution and strength of the Misner string singularities.  In particular, when $s=+n$, there is only one Misner string and it is located on the north pole ($\cos\theta=+1)$ axis, while the south pole $(\cos\theta=-1)$ axis is completely regular. The choice $s=-n$ corresponds to the opposite situation, while for $s=0$ the two Misner strings are `symmetrically distributed' and both axes are `equally singular'.

The parameter $s$ can be formally
eliminated by performing the  `large coordinate transformation' \cite{Clement:2015cxa}
\be
t\to t-{2s \phi}\,,
\ee
effectively threading the spacetime with an `overall Misner string' so that the symmetric distribution of Misner strings is restored. In general, changing the value of $s$ corresponds to a physical transformation of the system that simultaneously affects the strengths of both Misner strings. Such `variations of strings strengths' are captured by the first law of thermodynamics presented below. As shown in \cite{Clement:2015cxa} for non-rotating Taub--NUTs, the spacetime remains geodesically complete for any value of $s$, but the  absence of closed timelike and null geodesics requires  $|s/n|\leq 1$.

The spacetime is stationary and axisymmetric, corresponding to the $k=\partial_t$ and $\eta=\partial_\phi$ Killing vectors, and admits a number of Killing horizons.
The `black hole horizon' is located  at the largest root $r_+$ of $\Delta(r_+)=0$. It is generated by a (properly normalized at infinity) Killing vector
\be
\xi=k+\Omega \eta\,,
\ee
where $\Omega$ is the angular velocity of the ``Zero Angular Momentum Observer'' (ZAMO) evaluated at $r_+$:
\be\label{eq:OmegaH}
\Omega=-\frac{g_{t\phi}}{g_{\phi\phi}}\Big|_{r=r_+}=\frac{a}{r_+^2+a^2+n^2-2as}\,.
\ee
By standard arguments, it can be assigned the following entropy and temperature:
\ba
S&=&\frac{\mbox{Area}}{4}=\pi(r_+^2+a^2+n^2-2as)\,,\label{T}\\
T&=&\frac{\Delta'(r_+)}{4\pi(r_+^2+a^2+n^2-2as)}\,.\label{S}
\ea
Note that, contrary to the non-rotating case \cite{Bordo:2019tyh}, both quantities explicitly depend on the parameter
$s$.

As in the non-rotating case~\cite{Carlip:1999cy}, when Misner strings are present, the north and south pole axes are also Killing horizons, generated by the following Killing vectors:\footnote{Contrary to the Killing vector $\xi$ these cannot be properly normalized at infinity---their asymptotic norm depends on the angle $\theta$ and vanishes on the due part of the axis. The chosen normalization yields surface gravities that satisfy the first law below.}
\be
\xi_\pm=k- \frac{1}{2(s\pm n)} \eta\,.
\ee
The corresponding surface gravities $\kappa_\pm$ can be obtained from the standard formula $\kappa^2=-\frac{1}{2}(\nabla_\alpha \xi_\beta) (\nabla^\alpha \xi^\beta)$. Similar to \cite{Bordo:2019tyh} we associate with them the corresponding Misner potentials
\be\label{psipm}
\psi_\pm=\frac{\kappa_\pm}{4\pi}=\frac{1}{8\pi (n\pm s)}\,.
\ee

With these geometric quantities in hand, let us now turn towards thermodynamics.
Following \cite{Kubiznak:2019yiu, Bordo:2019tyh} we identify the temperature of the spacetime with the temperature of the black hole horizon, \eqref{T}, and the `entropy' of the spacetime with the entropy of the black hole horizon, \eqref{S}, and seek other thermodynamic charges so that the following generalized first law: \be\label{eq:FL}
\delta M=T \delta S+\Omega \delta J+ \psi_+ \delta N_+ +\psi_- \delta N_-\;,
\ee
together with the corresponding Smarr relation
\be\label{eq:Smarr}
M=2(TS+\Omega J+ \psi_+N_+ +\psi_-N_-)\;
\ee
are satisfied.
Obviously, such a first law if of full cohomogeneity: it has 4 terms on the r.h.s. which correspond to the variation of the 4 physical parameters of the solution: $\{r_+, a, n, s\}$.

To find the thermodynamic charges $M, J$, and $N_\pm$, let us generalize the Komar-like prescription developed in \cite{Bordo:2019tyh} to include rotation. The idea is to re-derive the Smarr formula, accounting properly for the new NUT-related boundaries of Misner tubes that surround Misner strings, and identify the due Komar integrals with the thermodynamic charges.

As any Killing vector, the generator of the black hole horizon obeys the following integrability condition:
\be\label{ids}
\nabla_a\nabla^a\xi^b=-R^b{}_a\xi^a=0\,,
\ee
with the last equality valid in vacuum spacetimes. Using the differential form language and integrating over the $3$-dimensional ($t=const.)$ hypersurface  $\Sigma$ we thus have
\be\label{integral}
0=\int_{\Sigma}d*d\xi
=\int_{\partial \Sigma}*d\xi\,.
\ee
The spacetime boundary $\partial \Sigma$ now consists of the two Misner string tubes $T_+$ and $T_-$ positioned at $\theta=\epsilon$ and $\theta=\pi-\epsilon$ respectively, the black hole horizon $H$ at $r=r_+$, and the sphere $S_\infty$ at $r=\infty$.
Accounting for the relative orientation of the boundaries it writes as
\be\label{eq:bdry}
\partial \Sigma= T_++S_{\infty}-T_--H\,,
\ee
see  \cite{Bordo:2019tyh} for more details.

We split the integral \eqref{integral} into the following parts:
\ba\label{eq:GemSmarr}
0&=&\int_{S_\infty}*dk-\int_{H}*d\xi  + \int_{\tiny T_+} \!\!*dk - \int_{\tiny T_-}\!\! *dk\nonumber\\
&&{+}\Omega\left(\int_{S^2_\infty}*d\eta+\int_{\tiny T_+}*d\eta-\int_{\tiny T_-}*d\eta\right)\;.
\ea
The first two terms are standard and simply yield the mass $M$ and the entropy $S$:
\ba
M&=&-\frac{1}{8\pi}\int_{S^2_\infty}*dk\,,\label{Mint}\\
S&=&-\frac{1}{16\pi T}\int_{H}*d\xi\,.\label{TSint}
\ea
The next two terms were already studied in \cite{Bordo:2019tyh}, and define the gravitational Misner charges:
\begin{align}\label{Nint}
N_\pm&\equiv\pm \frac{1}{16\pi \psi_\pm} \int_{\tiny T_\pm} \!\!*dk\,.
\end{align}
Finally, we define the total angular momentum $J$ as
\be\label{Jint}
J=\frac{1}{16\pi}\left(\int_{S^2_\infty}*d\eta+\int_{\tiny T_+}*d\eta-\int_{\tiny T_-}*d\eta\right)\,,
\ee
which obviously accounts for contributions of Misner strings. By construction, the defined quantities obey the generalized Smarr relation \eqref{eq:Smarr}.

It may at first seem strange to consider additional string contributions in the definition of the angular momentum. Let us note, however, that regarding the string itself a source of angular momentum has precedence in the literature~, e.g.,~\cite{Manko:2005nm, Aliev:2007fy, Aliev:2008wv}. Moreover, it is a somewhat remarkable fact that, individually, each of the three integrals in the definition of $J$ above is divergent, yet their sum produces a finite result.

Evaluating these integrals, we find the following thermodynamic quantities:
\ba
M&=&m\,,\\
N_\pm&=&-\frac{2\pi(s\pm n)^2(n\mp a)}{r_+}\,,\\
J&=&\frac{(r_+^2+a^2+n^2-2as)(a-s)}{2r_+}\,.\label{Jtot}
\ea
It is now easy to verify that these thermodynamic quantities, together with $S$ given by \eqref{S}, $T$ \eqref{T}, and $\psi$'s \eqref{psipm}, satisfy the first law \eqref{eq:FL}, providing a non-trivial consistency check of our results.

Thus, we have shown (for the first time ever) how to formulate consistent thermodynamics of rotating Taub--NUT solutions. The corresponding first law \eqref{eq:FL} is of full cohomogeneity and is valid for an arbitrary distribution of Misner strings. It also accounts for both: independent variations of the NUT parameter $n$ and independent variations of the Misner strength parameter $s$. As such, it may describe processes such as a capture of a rotating string by a black hole, or an axisymmetric merger of two Taub--NUT solutions.

To cross check our results, let us turn towards the Euclidean action and the corresponding derivation of the thermodynamic quantities. The action for our metric \eqref{NUT} can be obtained from the corresponding AdS action, see appendix, in the limit of the vanishing cosmological constant. It simply reads
\be\label{Gflat}
{\cal G}=\frac{m}{2}\,.
\ee
Calculated in the grandcanonical ensemble, it is a function of the corresponding quantities:
\be
\mathcal{G}=\mathcal{G}(T,\psi_+,\psi_-, \Omega)\,,
\ee
or more explicitly
\ba
\mathcal{G}&=&
\frac{T(\psi_+-\psi_-)}{16\Omega \psi_+\psi_-}-\frac{\pi T}{2 \Omega^2}\nonumber\\
&&+\frac{\sqrt{(4 \pi  \psi_++\Omega ) (4 \pi  \psi_--\Omega)(4\pi^2T^2+\Omega^2)}}{16\Omega^2\sqrt{\psi_+\psi_-}}\,.\qquad
\ea
It is then easy to verify that the thermodynamic quantities obtained above by the Komar integration can be recovered as
\be\label{jinak}
S=-\frac{\partial {\cal G}}{\partial T}\,,\quad N_\pm=-\frac{\partial {\cal G}}{\partial \psi_\pm}\,,\quad J=-\frac{\partial {\cal G}}{\partial \Omega}\,.
\ee
At the same time, the relation
\be
{\cal G}=M-TS-\Omega J-\psi_+N_+-\psi_-N_-
\ee
 is equivalent to the Smarr formula \eqref{eq:Smarr}.

Let us also note the following two interesting facts about the total angular momentum $J$. First, referring to the AdS case in the appendix again, one can employ the conformal method and calculate the asymptotic charge corresponding to the $\eta$ Killing vector, to find in the asymptotically flat limit
\be
J_0\equiv Q(\eta)=m(a-3s)\,.
\ee
This means that the total angular momentum $J$ actually differs from the asymptotic charge $J_0$ by the following Misner string contribution:
\be
J_s=J-J_0=\frac{a(s^2+n^2)+s(r_+^2-2n^2)}{r_+}\,.
\ee
In the special case where the strings are symmetrically distributed, then it is possible to compute finite values for $J_s$ and $J_0$ directly via the Komar integration---those values agree with the $s=0$ case above. Second, it is easy to see that as long as $s\neq 0$,  $J$ remains non-trivial also in the limit of zero rotation, $a\to 0$. However, there is no corresponding
term in the first law, as $\Omega$ in this limit automatically vanishes.

Let us conclude here with a few points for discussion. First, we emphasize that the first law we present should be interpreted as a first law of black hole \textit{mechanics}. Whether or not this first law corresponds, in the Lorentzian case, to a true thermodynamic first law for an equilibrium system depends on whether the potentials $\psi_\pm$ correspond to temperatures. If in fact they do, then the situation is somewhat analogous to de Sitter black holes, where one has multiple different temperatures and hence, in general, a non-equilibrium configuration. From this interpretation, regularity of the Euclidean sector---as considered in~\cite{Ghezelbash:2007kw}---is tantamount to enforcing equality of the temperatures and, therefore, equilibrium. However, unlike the case for black holes and cosmological horizons~\cite{Gibbons:1977mu}, there as yet exists no independent confirmation of whether or not $\psi_\pm$ corresponds to a genuine temperature.

{Our second point concerns the physical interpretation of $N_\pm$ provided $\psi_\pm$ indeed were temperatures.
The obtained first law clearly indicates that, being quantities conjugate to $\psi_\pm$, the charges $N_\pm$ would represent the associated thermodynamic entropy. However, similar to what has already been observed in the non-rotating case~\cite{Bordo:2019tyh}, $N_\pm$ are different from Noether charge entropy~\cite{Garfinkle:2000ms} one would naturally associate to the Killing horizons generated by $\xi_\pm$.  Namely, the Noether charge entropy $\hat N_\pm$ would be given by the integral of $\xi_\pm$ over the respective tube (normalized by surface gravity) rather than the integral of $k = \partial_t$ as presented above,
\be\label{Nother}
\hat N_\pm \equiv\pm \frac{1}{16\pi \psi_\pm} \int_{\tiny T_\pm} \!\!*d\xi_\pm\,.
\ee
The Noether prescription yields an (infinite) area of the tubes:
\be
\hat N_\pm =\frac{\mbox{Area}_\pm}{2}=2\pi (n\pm s)(R_c-r_+)\,
\ee
(where $R_c$ is the radial cutoff) and would have to be properly renormalized.  It remains an interesting open question as to why the first law involves a variation of only a `portion' of the Noether charge entropy encoded in $N_\pm$.
}


Lastly, let us note that a natural next step consists in extending the ideas presented here to the asymptotically AdS case. However, that situation is far subtler than one may naively expect. There are inherent issues associated with the thermodynamics of rotating AdS black holes having to do with, for example, the proper choice of time like Killing vector to produce to correct thermodynamic mass~\cite{Gibbons:2004ai}. The Kerr--Taub--NUT-AdS metrics present
these same---plus other---obstacles. We leave these issues for future investigations.

\section*{Acknowledgements}

This work was supported in part by the Natural Sciences and Engineering Research Council of Canada.
R.H. is grateful to the Banting
Postdoctoral Fellowship programme.
A.B., F.G., and D.K.\ acknowledge the Perimeter Institute for Theoretical Physics  for their support. Research at Perimeter Institute is supported by the Government of Canada through the Department of Innovation, Science and Economic Development Canada and by the Province of Ontario through the Ministry of Research, Innovation and Science.

\appendix
\section{Euclidean action}\label{AppFE}

To find the Euclidean action for the asymptotically flat solution \eqref{NUT} studied in the main text, we consider the rotating AdS Taub--NUT solution \cite{Griffiths:2005qp}, calculate its action by using the standard counterterms, and take the asymptotically flat limit.

The rotating AdS Taub--NUT solution reads
\ba\label{NUT2}
ds^2&=&-\frac{\Delta}{\Sigma}\Bigl[d\tau+(2n\cos \theta-a\sin^2\!\theta+2s)\frac{d\phi}{K}\Bigr]^2\\
&&+\frac{P}{\Sigma}\Bigl[ad\tau-(r^2+a^2+n^2-2as)\frac{d\phi}{K}\Bigr]^2\nonumber\\
&&+\frac{\Sigma}{\Delta}\,dr^2+\frac{\Sigma}{P}\sin^2\!\theta\, d\theta^2\,,
\ea
where
\ba
\Sigma&=&r^2+(n+a\cos\theta)^2\,,\\
\frac{P}{\sin^2\!\theta}&=&\,1-\frac{4an\cos\theta}{l^2} -\frac{a^2\cos^2\!\theta}{l^2}\,,\label{co}\\
\Delta&=&a^2\!-\!n^2\!-\!2mr\!+\!r^2\notag\\
&&+\frac{3(a^2\!-\!n^2)\,n^2+({a^2}\!+\!6n^2)\,r^2\!+\!{r^4}}{l^2}\;.\quad
\ea
Here, we have included the Misner string strength parameter $s$, $K$ is a constant to regulate the conical singularities on the axes, and $l$ is the AdS length.

The Euclidean action is calculated with the usual counter terms \cite{Emparan:1999pm}:
\ba\label{action}
I&=&\frac{1}{16\pi}\int_{M}d^{4}x\sqrt{g}\bigl( R+\frac{6}{ l^{2}}\bigr)\nonumber\\
&&+ \frac{1}{8\pi}\int_{\partial M}d^{3}x\sqrt{h}\left[\mathcal{K}
-   \frac{2}{ l} - \frac{ l}{2}\mathcal{R}\left( h\right) \right]\,,
\ea
where $\mathcal{K}$ and $\mathcal{R}\left( h\right)$ are
the extrinsic curvature and Ricci scalar of the boundary respectively. It is associated with a free energy, ${\cal G}=I/\beta$, where $\beta$ is the periodicity of the Euclidean time coordinate. This gives
\be\label{G}
{\cal G}_{\text{AdS}}=\frac{m}{2K}-\frac{r_+(a^2+r_+^2+3n^2)}{2Kl^2}\,.
\ee
The action \eqref{Gflat} in the main text is then obtained by  taking the asymptotically flat limit, $l\rightarrow\infty$, and setting $K=1$ (upon which the axes become regular for the corresponding choice of $s$).

\providecommand{\href}[2]{#2}\begingroup\raggedright\endgroup


\begin{thebibliography}{10}
	
	\bibitem{misner1963flatter}
	C.~W. Misner, \emph{The flatter regions of Newman, Unti, and Tamburino's
		generalized Schwarzschild space}, {\emph{Journal of Mathematical Physics}
		{\bfseries 4} (1963) 924}.
	
	\bibitem{bonnor1969new}
	W.~B. Bonnor, \emph{A new interpretation of the NUT metric in general
		relativity},  in \emph{Mathematical Proceedings of the Cambridge
		Philosophical Society}, vol.~66, pp.~145--151, Cambridge University Press,
	1969.
	
	\bibitem{Manko:2005nm}
	V.~S. Manko and E.~Ruiz, \emph{{Physical interpretation of NUT solution}},
	\href{https://doi.org/10.1088/0264-9381/22/17/014}{\emph{Class. Quant. Grav.}
		{\bfseries 22} (2005) 3555}
	[\href{https://arxiv.org/abs/gr-qc/0505001}{{\ttfamily gr-qc/0505001}}].
	
	\bibitem{miller1971taub}
	J.~Miller, M.~D. Kruskal and B.~B. Godfrey, \emph{Taub-NUT (Newman, Unti,
		Tamburino) metric and incompatible extensions}, {\emph{Physical Review D}
		{\bfseries 4} (1971) 2945}.
	
	\bibitem{Clement:2015cxa}
	G.~Cl\'{e}ment, D.~Gal'tsov and M.~Guenouche, \emph{{Rehabilitating space-times
			with NUTs}},
	\href{https://doi.org/10.1016/j.physletb.2015.09.074}{\emph{Phys. Lett.}
		{\bfseries B750} (2015) 591}
	[\href{https://arxiv.org/abs/1508.07622}{{\ttfamily 1508.07622}}].
	
	\bibitem{Clement:2015aka}
	G.~Cl\'{e}ment, D.~Gal'tsov and M.~Guenouche, \emph{{NUT wormholes}},
	\href{https://doi.org/10.1103/PhysRevD.93.024048}{\emph{Phys. Rev.}
		{\bfseries D93} (2016) 024048}
	[\href{https://arxiv.org/abs/1509.07854}{{\ttfamily 1509.07854}}].
	
	\bibitem{Clement:2016mll}
	G.~Cl\'{e}ment and M.~Guenouche, \emph{{Motion of charged particles in a NUTty
			Einstein-Maxwell spacetime and causality violation}},
	\href{https://doi.org/10.1007/s10714-018-2388-y}{\emph{Gen. Rel. Grav.}
		{\bfseries 50} (2018) 60} [\href{https://arxiv.org/abs/1606.08457}{{\ttfamily
			1606.08457}}].
	
	\bibitem{NouriZonoz:1998va}
	M.~Nouri-Zonoz and D.~Lynden-Bell, \emph{{Gravomagnetic lensing by NUT space}},
	{\emph{Mon. Not. Roy. Astron. Soc.} {\bfseries 292} (1997) 714}
	[\href{https://arxiv.org/abs/gr-qc/9812094}{{\ttfamily gr-qc/9812094}}].
	
	\bibitem{lynden1998classical}
	D.~Lynden-Bell and M.~Nouri-Zonoz, \emph{Classical monopoles: Newton, NUT
		space, gravomagnetic lensing, and atomic spectra}, {\emph{Reviews of Modern
			Physics} {\bfseries 70} (1998) 427}.
	
	\bibitem{Abdujabbarov:2012bn}
	A.~Abdujabbarov, F.~Atamurotov, Y.~Kucukakca, B.~Ahmedov and U.~Camci,
	\emph{{Shadow of Kerr-Taub-NUT black hole}},
	\href{https://doi.org/10.1007/s10509-012-1337-6}{\emph{Astrophys. Space Sci.}
		{\bfseries 344} (2013) 429}
	[\href{https://arxiv.org/abs/1212.4949}{{\ttfamily 1212.4949}}].
	
	\bibitem{Grenzebach:2014fha}
	A.~Grenzebach, V.~Perlick and C.~L\"{a}mmerzahl, \emph{{Photon Regions and Shadows
			of Kerr-Newman-NUT Black Holes with a Cosmological Constant}},
	\href{https://doi.org/10.1103/PhysRevD.89.124004}{\emph{Phys. Rev.}
		{\bfseries D89} (2014) 124004}
	[\href{https://arxiv.org/abs/1403.5234}{{\ttfamily 1403.5234}}].
	
	\bibitem{Goddi:2017pfy}
	C.~Goddi et~al., \emph{{BlackHoleCam: Fundamental physics of the galactic
			center}}, \href{https://doi.org/10.1142/S0218271817300014,
		10.1142/9789813226609_0046}{\emph{Int. J. Mod. Phys.} {\bfseries D26} (2016)
		1730001} [\href{https://arxiv.org/abs/1606.08879}{{\ttfamily 1606.08879}}].
	
	\bibitem{Akiyama:2019cqa}
	{\scshape Event Horizon Telescope} collaboration, K.~Akiyama et~al.,
	\emph{{First M87 Event Horizon Telescope Results. I. The Shadow of the
			Supermassive Black Hole}},
	\href{https://doi.org/10.3847/2041-8213/ab0ec7}{\emph{Astrophys. J.}
		{\bfseries 875} (2019) L1}.
	
	\bibitem{Hawking:1998ct}
	S.~W. Hawking, C.~J. Hunter and D.~N. Page, \emph{{NUT charge, anti-de Sitter
			space and entropy}},
	\href{https://doi.org/10.1103/PhysRevD.59.044033}{\emph{Phys. Rev.}
		{\bfseries D59} (1999) 044033}
	[\href{https://arxiv.org/abs/hep-th/9809035}{{\ttfamily hep-th/9809035}}].
	
	\bibitem{Chamblin:1998pz}
	A.~Chamblin, R.~Emparan, C.~V. Johnson and R.~C. Myers, \emph{{Large N phases,
			gravitational instantons and the nuts and bolts of AdS holography}},
	\href{https://doi.org/10.1103/PhysRevD.59.064010}{\emph{Phys. Rev.}
		{\bfseries D59} (1999) 064010}
	[\href{https://arxiv.org/abs/hep-th/9808177}{{\ttfamily hep-th/9808177}}].
	
	\bibitem{Emparan:1999pm}
	R.~Emparan, C.~V. Johnson and R.~C. Myers, \emph{{Surface terms as counterterms
			in the AdS/CFT correspondence}},
	\href{https://doi.org/10.1103/PhysRevD.60.104001}{\emph{Phys. Rev.}
		{\bfseries D60} (1999) 104001}
	[\href{https://arxiv.org/abs/hep-th/9903238}{{\ttfamily hep-th/9903238}}].
	
	\bibitem{Mann:1999pc}
	R.~B. Mann, \emph{{Misner string entropy}},
	\href{https://doi.org/10.1103/PhysRevD.60.104047}{\emph{Phys. Rev.}
		{\bfseries D60} (1999) 104047}
	[\href{https://arxiv.org/abs/hep-th/9903229}{{\ttfamily hep-th/9903229}}].
	
	\bibitem{Mann:1999bt}
	R.~B. Mann, \emph{{Entropy of rotating Misner string space-times}},
	\href{https://doi.org/10.1103/PhysRevD.61.084013}{\emph{Phys. Rev.}
		{\bfseries D61} (2000) 084013}
	[\href{https://arxiv.org/abs/hep-th/9904148}{{\ttfamily hep-th/9904148}}].
	
	\bibitem{Garfinkle:2000ms}
	D.~Garfinkle and R.~B. Mann, \emph{{Generalized entropy and Noether charge}},
	\href{https://doi.org/10.1088/0264-9381/17/16/314}{\emph{Class. Quant. Grav.}
		{\bfseries 17} (2000) 3317}
	[\href{https://arxiv.org/abs/gr-qc/0004056}{{\ttfamily gr-qc/0004056}}].
	
	\bibitem{Johnson:2014xza}
	C.~V. Johnson, \emph{{Thermodynamic Volumes for AdS-Taub-NUT and
			AdS-Taub-Bolt}},
	\href{https://doi.org/10.1088/0264-9381/31/23/235003}{\emph{Class. Quant.
			Grav.} {\bfseries 31} (2014) 235003}
	[\href{https://arxiv.org/abs/1405.5941}{{\ttfamily 1405.5941}}].
	
	\bibitem{Johnson:2014pwa}
	C.~V. Johnson, \emph{{The Extended Thermodynamic Phase Structure of Taub-NUT
			and Taub-Bolt}},
	\href{https://doi.org/10.1088/0264-9381/31/22/225005}{\emph{Class. Quant.
			Grav.} {\bfseries 31} (2014) 225005}
	[\href{https://arxiv.org/abs/1406.4533}{{\ttfamily 1406.4533}}].
	
	\bibitem{Astefanesei:2004ji}
	D.~Astefanesei, R.~B. Mann and E.~Radu, \emph{{Breakdown of the entropy/area
			relationship for NUT-charged spacetimes}},
	\href{https://doi.org/10.1016/j.physletb.2005.05.057}{\emph{Phys. Lett.}
		{\bfseries B620} (2005) 1}
	[\href{https://arxiv.org/abs/hep-th/0406050}{{\ttfamily hep-th/0406050}}].
	
	\bibitem{Mann:2004mi}
	R.~B. Mann and C.~Stelea, \emph{{On the thermodynamics of NUT charged spaces}},
	\href{https://doi.org/10.1103/PhysRevD.72.084032}{\emph{Phys. Rev.}
		{\bfseries D72} (2005) 084032}
	[\href{https://arxiv.org/abs/hep-th/0408234}{{\ttfamily hep-th/0408234}}].
	
	\bibitem{Holzegel:2006gn}
	G.~Holzegel, \emph{{A Note on the instability of Lorentzian Taub-NUT-space}},
	\href{https://doi.org/10.1088/0264-9381/23/11/017}{\emph{Class. Quant. Grav.}
		{\bfseries 23} (2006) 3951}
	[\href{https://arxiv.org/abs/gr-qc/0602045}{{\ttfamily gr-qc/0602045}}].
	
	\bibitem{Kubiznak:2019yiu}
	R.~A. Hennigar, D.~Kubiznak and R.~B. Mann, \emph{{Thermodynamics of Lorentzian
			Taub-NUT spacetimes}},  \href{https://arxiv.org/abs/1903.08668}{{\ttfamily
			1903.08668}}.
	
	\bibitem{Ballon:2019uha}
	A.~B. Bordo, F.~Gray and D.~Kubiznak, \emph{{Thermodynamics and Phase
			Transitions of NUTty Dyons}},
	\href{https://arxiv.org/abs/1904.00030}{{\ttfamily 1904.00030}}.
	
	\bibitem{Bueno:2018uoy}
	P.~Bueno, P.~A. Cano, R.~A. Hennigar and R.~B. Mann, \emph{{NUTs and bolts
			beyond Lovelock}}, \href{https://doi.org/10.1007/JHEP10(2018)095}{\emph{JHEP}
		{\bfseries 10} (2018) 095}
	[\href{https://arxiv.org/abs/1808.01671}{{\ttfamily 1808.01671}}].
	
	\bibitem{Ghezelbash:2007kw}
	A.~M. Ghezelbash, R.~B. Mann and R.~D. Sorkin, \emph{{The Disjointed
			thermodynamics of rotating black holes with a NUT twist}},
	\href{https://doi.org/10.1016/j.nuclphysb.2007.04.004}{\emph{Nucl. Phys.}
		{\bfseries B775} (2007) 95}
	[\href{https://arxiv.org/abs/hep-th/0703030}{{\ttfamily hep-th/0703030}}].
	
	\bibitem{Bordo:2019tyh}
	A.~B. Bordo, F.~Gray, R.~A. Hennigar and D.~Kubiznak, \emph{{Misner
			Gravitational Charges and Variable String Strengths}},
	\href{https://arxiv.org/abs/1905.03785}{{\ttfamily 1905.03785}}.
	
	\bibitem{Carlip:1999cy}
	S.~Carlip, \emph{{Entropy from conformal field theory at Killing horizons}},
	\href{https://doi.org/10.1088/0264-9381/16/10/322}{\emph{Class. Quant. Grav.}
		{\bfseries 16} (1999) 3327}
	[\href{https://arxiv.org/abs/gr-qc/9906126}{{\ttfamily gr-qc/9906126}}].
	
	\bibitem{Aliev:2007fy}
	A.~N. Aliev, \emph{{Rotating Spacetimes with Asymptotic Non-Flat Structure and
			the Gyromagnetic Ratio}},
	\href{https://doi.org/10.1103/PhysRevD.77.044038}{\emph{Phys. Rev.}
		{\bfseries D77} (2008) 044038}
	[\href{https://arxiv.org/abs/0711.4614}{{\ttfamily 0711.4614}}].
	
	\bibitem{Aliev:2008wv}
	A.~N. Aliev, H.~Cebeci and T.~Dereli, \emph{{Kerr-Taub-NUT Spacetime with
			Maxwell and Dilaton Fields}},
	\href{https://doi.org/10.1103/PhysRevD.77.124022}{\emph{Phys. Rev.}
		{\bfseries D77} (2008) 124022}
	[\href{https://arxiv.org/abs/0803.2518}{{\ttfamily 0803.2518}}].
	
	\bibitem{Gibbons:1977mu}
	G.~W. Gibbons and S.~W. Hawking, \emph{{Cosmological Event Horizons,
			Thermodynamics, and Particle Creation}},
	\href{https://doi.org/10.1103/PhysRevD.15.2738}{\emph{Phys. Rev.} {\bfseries
			D15} (1977) 2738}.
	
	\bibitem{Gibbons:2004ai}
	G.~W. Gibbons, M.~J. Perry and C.~N. Pope, \emph{{The First law of
			thermodynamics for Kerr-anti-de Sitter black holes}},
	\href{https://doi.org/10.1088/0264-9381/22/9/002}{\emph{Class. Quant. Grav.}
		{\bfseries 22} (2005) 1503}
	[\href{https://arxiv.org/abs/hep-th/0408217}{{\ttfamily hep-th/0408217}}].
	
	\bibitem{Griffiths:2005qp}
	J.~B. Griffiths and J.~Podolsky, \emph{{A New look at the Plebanski-Demianski
			family of solutions}},
	\href{https://doi.org/10.1142/S0218271806007742}{\emph{Int. J. Mod. Phys.}
		{\bfseries D15} (2006) 335}
	[\href{https://arxiv.org/abs/gr-qc/0511091}{{\ttfamily gr-qc/0511091}}].
	
\end{thebibliography}
\end{document}